\documentstyle[12pt,epsfig]{article}

\def \sect #1 {\setcounter{equation} 0\section{#1}}
\def \be  {\begin{equation}}
\def \ee  {\end{equation}}
\def \ba  {\begin{eqnarray}}
\def \ea  {\end{eqnarray}}
\def \baa {\begin{eqnarray*}}
\def \eaa {\end{eqnarray*}}
\def \bb  {\begin{thebibliography}}
\def \eb  {\end{thebibliography}}

\def \CO {{\cal O}}

\def \lab #1 {\label{#1}}

\def \CO {{\cal O}}

\def \fracs #1#2 {\mbox{\small $\frac{#1}{#2}$}}

\def \bin #1#2 {{\left({#1}\atop{#2}\right)}}
\def\lapproxeq{{\ \lower 0.6ex \hbox{$\buildrel<\over\sim$}\ }}
\def\gapproxeq{{\ \lower 0.6ex \hbox{$\buildrel>\over\sim$}\ }}

\def \as {\relax\ifmmode\alpha_s\else{$\alpha_s${ }}\fi}

\def \al #1 {\frac {\as({#1})}{\pi} }
\def \ds #1 {\ooalign{$\hfil/\hfil$\crcr$#1$}}

\def \GeV {\mbox{GeV}}
\def \prt{perturbative }
\def \nprt{nonperturbative }

\def \CO {{\cal O}}
\def\np#1#2#3  {{Nucl. Phys.~{\bf #1} (19#3) #2}}
\def\nc#1#2#3  {{Nuovo. Cim.~{\bf #1} (19#3) #2}}
\def\pl#1#2#3  {{Phys. Lett.~{\bf #1} (19#3) #2}}
\def\pr#1#2#3  {{Phys. Rev.~{\bf #1} (19#3) #2}}
\def\prl#1#2#3  {{Phys. Rev. Lett.~{\bf #1} (19#3) #2}}
\def\prep#1#2#3 {{Phys. Rep.~{\bf #1} (19#3) #2}}
\def\zp#1#2#3  {{Z. Phys.~{\bf #1} (19#3) #2}}
\def\epj#1#2#3  {{Eur. Phys. J.~{\bf #1} (#3) #2}}
\def\rmp#1#2#3  {{Rev. Mod. Phys.~{\bf #1} (19#3) #2}}
\def\JETP#1#2#3 {{Sov.\ Phys.\ JETP~{\bf #1} (19#3) #2}}
\def\sj#1#2#3 {{Sov.\ J.\ Nucl.\ Phys.~{\bf #1} (19#3) #2}}
\def\hepph  #1 {{hep-ph/#1}}

\begin{document}

\begin{flushright}
BNL-HET-01/43 \\
BNL-NT-01/32 \\
RBRC-232 \\
YITP-SB-01-75\\
\today
\end{flushright}

\vspace*{30mm}

\begin{center}
{\LARGE Joint resummation in }\\

\vskip 5mm

{\LARGE electroweak boson production}

\par\vspace*{20mm}\par

{\large Anna Kulesza$^a$,
George Sterman$^b$, Werner Vogelsang$^c$}

\bigskip

{\em $^a$Department of Physics, Brookhaven National Laboratory,
Upton, NY 11973, U.S.A.}

\bigskip

{\em $^b$C.N.\ Yang Institute for Theoretical Physics,
SUNY Stony Brook\\
Stony Brook, New York 11794 -- 3840, U.S.A.}

\bigskip

{\em $^c$RIKEN-BNL Research Center and Nuclear Theory, \\
Brookhaven National Laboratory,
Upton, NY 11973, U.S.A.}

\end{center}
\vspace*{15mm}

\begin{abstract}
\noindent
We present a phenomenological application of the joint resummation
formalism to electroweak annihilation processes at measured
boson momentum $Q_T$. This formalism
simultaneously resums at next-to-leading logarithmic accuracy
large threshold and recoil corrections to partonic scattering. 
We invert the impact parameter transform using a previously described 
analytic continuation procedure.  This leads to a well-defined,
resummed perturbative cross section for all nonzero $Q_T$, which can be 
compared to resummation carried out directly in $Q_T$ space. From 
the structure of the resummed expressions, we 
also determine the
form of nonperturbative corrections to the cross section
and implement these into our analysis.
We obtain
a good description of the transverse momentum distribution of
Z bosons produced at the Tevatron collider.
\end{abstract}

\newpage

\section{Introduction}

The hadronic annihilation
cross sections for electroweak
boson production ($\gamma^*$, W, Z, H)
provide important test cases for resummation techniques
in QCD.  This paper describes the application
of a  simultaneous, joint
resummation of threshold and transverse momentum
singularities in these cross sections.
The application
to this well-studied case enables us to
compare joint resummation to the data
for Z production at the Tevatron.
We regard it as a first step toward extending
joint resummation to a wider range in
kinematics and reactions, and toward a unified
description of nonperturbative contributions
in hadronic reactions.

At  measured transverse momentum $Q_T\ll Q$,
with $Q$ the pair or boson mass,
annihilation cross sections have
contributions $\alpha_s^n\ln^{2n-1}(Q_T^2/Q^2)/Q_T^2$
at each order of perturbation theory.
Although highly singular at $Q_T=0$,
these terms nevertheless
organize themselves into a function that
describes the Sudakov suppression of the cross
section at small $Q_T$ \cite{ktorig,AEGM,Coll82,Coll85}.  These singularities
reflect the recoil of the produced electroweak boson
against soft gluon radiation.   This is transverse
momentum resummation, or ``$Q_T$" resummation,
for short.

Perturbative corrections
in factorized inclusive
electroweak annihilation cross sections,
\be
\label{inclusive}
     \frac{d\sigma_{AB}^{\rm res}}{dQ^2}
     =  \int dz \int dx_i\, dx_j\; f_{i/A} (x_i,\mu_F)
\; f_{j/B} (x_j ,\mu_F)\; \delta\left(z-Q^2/x_ix_jS\right)\;
\; {\omega}_{ij}^{\rm res} (z,Q,\mu,\mu_F) \; ,
\ee
have an analogous behavior at $z=1$.  In this expression,
the $f$'s are parton distribution functions
in hadrons $A$ and $B$, with $\omega_{ij}^{\rm res}$
the corresponding partonic hard-scattering functions, $\mu$ is
the renormalization scale and $\mu_F$ the factorization scale.
At $z=1$, referred to as partonic threshold,
$\omega_{ij}^{\rm res}(z)$ has terms of the form
$\alpha_s^n\ln^{2n-1}(1-z)/(1-z)$.
These singular corrections reflect
the lack of phase space for soft radiation
when the partons have just enough energy to
produce the observed final state.
Like singularities at $Q_T=0$, they can be resummed to all
orders \cite{Ste87,CT8991}.

Unlike transverse
momentum resummation, the singular functions of
threshold resummation do not appear as explicit
logarithms in the physical cross section, Eq.\ (\ref{inclusive}),
since they are in convolution with
the parton distribution functions. In practice,
transverse momentum resummation
is usually of greater numerical significance
at measured $Q_T$ than is threshold resummation
at integrated $Q_T$,
simply because the typical value of $z$ in
Eq.\ (\ref{inclusive}) is generally far from unity.
In such cases, threshold resummation
shows that the effects of distributions singular
at $z=1$ are under control.   It also leads to
cross sections that are in general less sensitive
to the factorization scale \cite{Ber98a}.   For this reason, and because
the dynamical origins of the large corrections in
both threshold and transverse momentum resummations
are in soft gluon emission, it  is attractive to
develop a formalism that encompasses both.  The
necessary analysis for this combination, which
we refer to as joint resummation, has been carried out
in \cite{Li99} for parton distributions,
and in \cite{lsv} for electroweak annihilation as
well as QCD cross sections.
The effects of resummation are closely bound to momentum
conservation.  The singular corrections associated with
soft gluon emission exponentiate in the corresponding spaces of moments,
impact parameter for transverse momentum, and Mellin
(or Laplace) moments in energy for
threshold resummation.   The transforms relax momentum and energy
conservation, while their inverses reimpose it.  In
joint resummation, both transverse momentum and energy conservation
are respected.

In Ref.\ \cite{lsvprl}, a preliminary
analysis of prompt photon production was carried out
from the point of view of joint resummation, with an
emphasis on the role of recoil in higher-order
terms in the relevant hard scattering functions.
Relatively substantial effects were found
in that case.  In this context, it is important to return to
the better-understood example of electroweak annihilation
to test the joint formalism.

As mentioned above, there is a close
relation between resummation and nonperturbative
corrections.  Taking into account
nonleading logarithms and
the running of the coupling, resummation leads
in each case to a perturbative
expression in  which
the scale of the coupling reflects the
value of the transform variable.
Because of the singularity of
the perturbative effective coupling at $\Lambda_{\rm QCD}$,
the resulting expressions are, strictly speaking,
undefined.  A closer look, however, shows that
singular contributions appear only
at nonleading powers of momentum transfer. This is an
example of how perturbative resummation can
suggest the way nonperturbative
dynamics is expressed in infrared safe
hard scattering functions.   In effect, perturbation theory
is ambiguous, and the resolution of its ambiguities is, by
definition, nonperturbative \cite{cspv,irr}.   As we shall
review below, these ambiguities manifest themselves as
singularities in the integrand of the
inverse transforms for both
transverse momentum and threshold resummations.
Each scheme for dealing with these
singularities constitutes a specification of perturbation theory, and
implies a
parameterization of nonperturbative effects.   We hope that a
joint resummation affords a more general approach to this
problem.

For each resummed cross section, be it $Q_T$, threshold or
joint, one must first specify how to define
perturbation theory in the transform space,
and then, given a well-defined expression,
how to invert the transform.  The most ``conservative"
approach is to expand the resummed expression to
some fixed order in perturbation theory \cite{KIDfiniteorder}.  Since
the resummation contains information on singular
terms at all orders, we may in this way
get information beyond the order at which a
complete calculation has been done.
To fixed order, the perturbative expression and
its transform are unambiguous.
This approach is  applicable to threshold
resummation for inclusive cross sections,
but is not directly useful for phenomenology at measured $Q_T\sim 0$,
where the cross section is singular.
Another approach for $Q_T$ resummation is to use the
exponentiation of leading logarithms in
transverse momentum space, and incorporate
the finite order improvements, including those
due to the running coupling and momentum conservation, by a direct expansion
\cite{ERV,KS}. A similar application for threshold resummation
was developed in \cite{cs}.

Finally, in both $Q_T$ and threshold resummation,
we may, as indicated above,
redefine the resummed perturbation theory in
transform space, and invert the resulting transform
numerically.  For threshold resummation this has been
done by so-called principal-value \cite{cspv,Ber98} and minimal \cite{cmnt}
prescriptions for perturbation theory.  Both
exploit the analytic structure of the running
coupling, and redefine transform integrals
to avoid the Landau pole.
For $Q_T$ resummation, the commonly-used
approach (the $b_*$-prescription) \cite{Coll85,ds,AK,ly,Res}  introduces
an infrared scale, beyond which the running of
the coupling is decoupled from the transform
variable.  More recently, Qiu and Zhang \cite{Qiu00} have
proposed another method, employing
a smooth extrapolation of the perturbative resummed
cross section in impact parameter space
to push the Landau singularity to infinity.

In this paper, we will study the jointly resummed
cross section derived in \cite{lsv},
defined by a next-to-leading logarithm approximation
to the resummed exponent, and an
analytic continuation of the $b$-space contour
 following
\cite{lsvprl}.  We stress that we could have
made other choices, like those mentioned above,
within the context of joint resummation.  We
find the approach of \cite{lsvprl},
reviewed below, attractive for
its simplicity.  It results in a redefined
perturbative series with no new dimensional
scale beyond $\Lambda_{\rm QCD}$.  Phenomenological
parameters then appear only as nonperturbative
power corrections.  In this ``test case" for
joint resummation, we shall find a
consistency with the data from Z production that
we believe is comparable to that of other
approaches.  We hope that this will serve as
a starting-point for further development.

We note that different treatments of the
perturbative exponent may differ markedly
in pure threshold resummation, due to
different treatments of nonleading 
logarithms~\cite{Ber98,cmnt,Nick01}.  
This important question
applies to joint resummation as well, but
we shall not explore these differences
in this paper.  This is because
the vector boson cross sections discussed here
are in the kinematic region described above,
where $z$ is far from unity in Eq.\ (\ref{inclusive}), and
where we expect the effect of pure threshold resummation to be
small in any scheme. 

We begin our technical discussion with a review of the joint resummation
formalism of Ref.\ \cite{lsv}, applied to Z production, including
a comparison to the exact ${\cal O}(\as)$ cross section. 
We develop the phenomenological evaluation
of the resulting expressions in Sec.\ 3.
This includes
the specification of resummed perturbation theory
through contour integrals in the two transform spaces, the
matching to finite order calculations, and the
parameterization of nonperturbative effects.
In Sec.\ 4, we compare the resulting expressions
to CDF and D0 data, fitting the necessary nonperturbative
parameter.  Following this, we give a few
preliminary conclusions on the status and prospects
of the joint resummation program.  In Appendix A,
we present some well-known formulas that are relevant
for our calculation, and in Appendix B,
we sketch the simple numerical method that enables us to evaluate
resummed moment space integrals for parton distribution
functions that are specified only in $x$-space.

\section{The jointly resummed cross section}

Within the formalism of~\cite{lsv}, the jointly resummed cross section
$d\sigma^{\rm res}/dQ^2\,dQ_T^2$ for electroweak annihilation
is obtained as a double inverse Mellin and Fourier (impact parameter)
transform:
\ba
\label{crsec}
     \frac{d\sigma_{AB}^{\rm res}}{dQ^2\,dQ_T^2}
     &=&    \sum_a \sigma_{a}^{(0)}\,
\int_{C_N}\, \frac{dN}{2\pi i} \,\tau^{-N}\;    \int \frac{d^2b}{(2\pi )^2} \,
e^{i{\vec{Q}_T}\cdot {\vec{b}}}\, \nonumber \\
&\times&    {\cal C}_{a/A}(Q,b,N,\mu,\mu_F )\;
      \exp\left[ \,E_{a\bar a}^{\rm PT} (N,b,Q,\mu)\,\right] \;
      {\cal C}_{\bar{a}/B}(Q,b,N,\mu,\mu_F) \; .
\ea
Here, $\tau=Q^2/S$, and 
$\sigma_{a}^{(0)}$ is a normalization containing the
appropriate electroweak charges occurring in the basic underlying
process $a\bar{a}\to V$. For completeness, and because this is 
the process we will study numerically in this paper, we give
$\sigma_{a}^{(0)}$ for Z production in Appendix~A.

Compared to the expression given in~\cite{lsv},
we have brought the cross section into a form
that is closer to the standard one~\cite{Coll85}
in $Q_T$ resummation. Resummation of
soft-gluon effects is achieved through the flavor-diagonal
Sudakov exponent $E_{a\bar a}^{\rm PT}
(N,b,Q,\mu)$, while the ``${\cal C}$-coefficients'' contain the
parton distribution functions and provide
resummation of additional logarithms of soft-collinear and collinear origin. 
We will discuss the various terms in turn.

\subsection{The Sudakov exponent}

We now develop an expression for the exponent $E_{a\bar a}^{\rm PT}$,
valid to next-to-leading logarithms (NLL)
in $N$ and $b$, based on the results of Ref.~\cite{lsv}. 
Compared to that reference, we will absorb some terms in the exponent 
that are associated with  parton evolution into the ${\cal C}$-coefficients. 
We begin with the exponent derived in~\cite{lsv} for the eikonal
approximation to $a\bar a$ annihilation to an electroweak boson,
$E_{a\bar a}^{\rm eik}$. It is given by:
\ba
E_{a\bar a}^{\rm eik} (N,b,Q,\mu,\mu_F)
&=& 2 \int_0^{Q^2} {d k_T^2 \over
k_T^2} \;
A_a(\as(k_T))\left[J_0(b k_T) \, K_0\left({2 N k_T \over Q}\right) +
\ln\left({\bar N k_T \over Q}\right)\right]  \nonumber \\
&&-\,2 \ln\left(\bar{N}\right) \int_{\mu_F^2}^{Q^2} {d k_T^2 \over
k_T^2} A_a(\as(k_T)) \; .
\label{eikexpdef}
\ea
Here, $J_0$ and $K_0$ are the usual Bessel functions, and we define
\be
\label{nbardef}
\bar{N} = N{\rm e^{\gamma_E}} \; ,
\ee
with $\gamma_E$ the Euler constant.
The function  $A_a(\as)$ is a series in $\as$,
\be
A_a(\as) = \frac{\as}{\pi} A_a^{(1)} +  \left( \frac{\as}{\pi}\right)^2
A_a^{(2)} + \ldots \; ,
\ee
in terms the familiar coefficients~\cite{a2calc}:
\ba
A_a^{(1)} &=&  C_a \;\;\;\;\;\;\;\;\; (C_q=C_F, \; C_g =C_A)\;\; ,\nonumber \\
A_a^{(2)} &=& \frac{C_a}{2} K \;\;\;\; , \;\;\;\;\;
K=C_A \left( \frac{67}{18}-\frac{\pi^2}{6} \right) -\frac{10}{9}T_R N_F\; .
\ea
Dependence on the renormalization scale is implicit in Eq.~(\ref{eikexpdef})
through the expansion of $\as(k_T)$ in powers of $\as(\mu)$.

Following Refs.~\cite{lsv,Coll85,ctmin}, we approximate 
the exponent in Eq.\ (\ref{eikexpdef}) by a ``minimal" form
that is accurate to next-to-leading logarithm in both transform variables:
\ba
E_{a\bar a}^{\rm eik} (N,b,Q,\mu,\mu_F) &=&
2 \int_{Q^2/\chi^2}^{Q^2} {d k_T^2 \over k_T^2} \;A_a(\as(k_T))\,
\ln\left( {\bar N k_T\over Q} \right)-\,2 \ln\left(\bar{N}\right)
\int_{\mu_F^2}^{Q^2} {d k_T^2 \over k_T^2} A_a(\as(k_T)) \; .
\nonumber\\
&\ &
\label{Elog}
\ea
Here, the function $\chi(\bar{N},\bar{b})$ 
organizes the logarithms of $N$ and $b$ in joint resummation,
\be
\label{chinew}
\chi(\bar{N},\bar{b})=\bar{b} + \frac{\bar{N}}{1+\eta\,\bar{b}/
\bar{N}}\; ,
\ee
where $\eta$ is a constant, and
where, by analogy to $\bar{N}$ of Eq.~(\ref{nbardef}), we have 
defined 
\be
\label{bbardef}
\bar{b}\equiv b Q {\rm e^{\gamma_E}}/2 \; .
\ee
To anticipate, we will
choose $\eta=1/4$ below.
We will discuss the particular expression (\ref{chinew}) 
for $\chi$ 
in the next subsection.

We now regroup Eq.~(\ref{Elog}) as
\ba
\!\!\!
E_{a\bar a}^{\rm eik} (N,b,Q,\mu,\mu_F) &=&
2 \int_{Q^2/\chi^2}^{Q^2} {d k_T^2 \over k_T^2} \;A_a(\as(k_T))\,
\ln\left( {k_T\over Q} \right)-\,2 \ln\left(\bar{N}\right)
\int_{\mu_F^2}^{Q^2/\chi^2} {d k_T^2 \over k_T^2} A_a(\as(k_T)) \; .
\nonumber\\
&\ &
\label{Elog1}
\ea
Next, we make contact with
standard $Q_T$ resummation by writing
\ba
E_{a\bar a}^{\rm eik} (N,b,Q,\mu,\mu_F) &=&
-\int_{Q^2/\chi^2}^{Q^2} {d k_T^2 \over k_T^2} \;
\left[ A_a(\as(k_T))\,
\ln\left( {Q^2 \over k_T^2} \right) + B_a(\as(k_T))\right] \nonumber \\
&+&\int_{\mu_F^2}^{Q^2/\chi^2} {d k_T^2 \over k_T^2}
\Big[ -2 A_a(\as(k_T)) \ln\left( \bar{N}\right)
- B_a(\as(k_T)) \Big] \; .
\label{Elog2}
\ea
Here we have introduced
\be
B_a(\as) = \frac{\as}{\pi} B_a^{(1)} +  \left( \frac{\as}{\pi}\right)^2
B_a^{(2)} + \ldots \; ,
\ee
with
\ba
B_q^{(1)} &=&-\frac{3}{2} C_F \;\;\;\; , \;\;\;\;\;
B_g^{(1)} \;=\;-\frac{1}{6} \left( 11 C_A - 4 T_R N_F \right) \; .
\ea
Eq.~(\ref{Elog2}) follows from Eq.~(\ref{Elog1}) to NLL accuracy
in $N$ and $b$.
The coefficients $B_a^{(2)}$ are also known~\cite{ds,daniel}, but
contribute only beyond NLL. 

The first term in Eq.~(\ref{Elog2}),
\be \label{choice}
-\int_{Q^2/\chi^2}^{Q^2} {d k_T^2 \over k_T^2} \;
\left[ A_a(\as(k_T))\,
\ln\left( {Q^2 \over k_T^2} \right) + B_a(\as(k_T))\right]  \; ,
\ee
has the classic form of the Sudakov exponent in electroweak annihilation, 
the only new ingredient being the quantity $\chi$ that depends on $N$ and
$b$ and represents the joint resummation. As shown in Sec.~3.1 of 
Ref.~\cite{lsv}, the term with $B_a$ accounts for the difference 
between the eikonal approximation and the full partonic cross sections 
in the threshold region. Eq.~(\ref{choice}) will be
our choice for the exponent $E_{a\bar a}^{\rm PT}$ in Eq.~(\ref{crsec}).
Its expansion in leading and next-to-leading logarithms gives
\begin{equation}
\label{expdef}
E_{a\bar a}^{\rm PT} (N,b,Q,\mu) = \frac{2}{\alpha_s (\mu)}
h_a^{(0)} (\beta) +
2h_a^{(1)} (\beta,Q,\mu)   \;  ,
\end{equation}
where
\begin{eqnarray}
h_a^{(0)} (\beta) &=& \frac{A_a^{(1)}}{2\pi b_0^2}
\left[ 2 \beta + \ln(1-2 \beta) \right]\, ,\\
h_a^{(1)} (\beta,Q,\mu) &=&
\frac{A_a^{(1)} b_1}{2\pi b_0^3} \left[ \frac{1}{2} \ln^2 (1-2 \beta) +
\frac{2 \beta + \ln(1-2 \beta)}{1-2\beta} \right] + 
\frac{B_a^{(1)}}{2\pi b_0}  \ln(1-2 \beta) \nonumber \\
&+& \frac{1}{2\pi b_0} \left[ A_a^{(1)}\ln \left( \frac{Q^2}{\mu^2} \right)
-\frac{A_a^{(2)}}{\pi b_0}\right] \;
\left[ \frac{2 \beta}{1-2\beta}+ \ln(1-2 \beta) \right] \; .
\label{hsubadef}
\end{eqnarray}
In these equations,
\begin{eqnarray}
\lambda &=& b_0\, \alpha_s (\mu)  \ln \left (\bar{N}  \right) \; ,
\nonumber \\
\beta &=& b_0\, \alpha_s (\mu)
\ln \left( \chi \right) \, ,
\label{varsdef}
\ea
and
\ba
b_0 &=& \frac{11 C_A - 4 T_R N_F}{12 \pi}\;\;\;\; , \;\;\;\;\;
b_1 \;=\; \frac{17 C_A^2-10 C_A T_R N_F-6 C_F T_R N_F}{24 \pi^2}\; .
\end{eqnarray}

In order to interpret the second term in Eq.~(\ref{Elog2}), we note 
that the combination in its square brackets,
\be
\label{leadN}
-2 A_a(\as) \ln\left( \bar{N}\right)
- B_a(\as) \; ,
\ee
corresponds to the leading (logarithmic plus constant) terms at 
large $N$ in the one-loop diagonal $q\to q$ (or, $g\to g$) anomalous 
dimension. To be more precise, the way we have obtained
Eq.~(\ref{Elog2}), the second term in Eq.~(\ref{Elog2}) matches
the anomalous dimension to NLL accuracy only,
since the constant (in $N$) part of the two-loop anomalous
dimension is not identical to the customary~\cite{ds,AK,ly,Res}
coefficient $B_a^{(2)}$
(even though it is related to it~\cite{daniel,CDG}). As mentioned above, 
contributions related to $B_a^{(2)}$ enter only beyond NLL and are outside
the presently developed reach of the joint resummation formalism. We are
therefore indeed free to associate the combination in Eq.~(\ref{leadN})
with the leading terms in the anomalous dimension.  It then follows that
the second term in Eq.~(\ref{Elog2}) represents the evolution of the 
parton densities from scale $\mu_F$ to scale $Q/\chi$ in the large-$N$ 
limit, that is, near threshold. Note that indeed all dependence
of the exponent $E_{a\bar a}^{\rm eik}$ on $\mu_F$ is contained
in this term. The evolution term we have identified 
in Eq.~(\ref{Elog2}) will become part 
of the functions ${\cal C}_{a/H}$ introduced in Eq.~(\ref{crsec}).
We shall therefore pursue it further when discussing the ${\cal C}_{a/H}$ in 
subsection 2.3. below. 
\subsection{The function $\chi(\bar{N},\bar{b})$}
We have defined $\chi(\bar{N},\bar{b})$ 
in Eq.~(\ref{chinew}). There is an element of choice
in the actual form of $\chi(\bar{N},\bar{b})$, the only requirement
being that the leading and next-to-leading logarithms of $\bar{N}$ and $\bar b$ are
correctly reproduced in the limits $\bar{N}\to \infty$ or $\bar{b}\to\infty$,
respectively. In ref.~\cite{lsv}, the somewhat simpler choice
\be
\label{chiold}
\chi (\mbox{Ref.~\cite{lsv}})= \bar{b} +\bar N
\ee
was made. While this is a legitimate option, we found it to be
less convenient for phenomenological studies. The reason for this is
that this form of $\chi$ introduces sizable subleading
terms into perturbative expansions of the resummed exponent, which are
not present in full fixed-order perturbative results. For instance, expanding
the exponent $E_{a\bar a}^{\rm PT} (N,b,Q,\mu)$ in Eq.~(\ref{expdef}) 
to ${\cal O}(\alpha_s(\mu))$ one finds:
\be
\label{expb}
\exp\left[ \,E_{a\bar a}^{\rm PT} (N,b,Q,\mu) \,\right]
\;\approx\; 1 - \frac{2 \alpha_s(\mu)}{\pi} \,C_F\,\left[
\ln^2 (\chi) -\frac{3}{2} \ln (\chi) \right]\; .
\ee
If we are assuming that $\bar{b}\gg \bar{N}$, relevant at small $Q_T$
far away from threshold, then this is approximately
\be
\label{approb}
\exp\left[ \,E_{a\bar a}^{\rm PT} (N,b,Q,\mu) \,\right]
\;\approx\; 1 - \frac{2 \alpha_s(\mu)}{\pi} \,C_F\,\left[
\ln^2 (\bar{b}) -\frac{3}{2}\ln (\bar{b})
+ 2 \frac{\bar{N}}{\bar{b}} \ln (\bar{b})+
{\cal O}\left( \frac{1}{\bar{b}} \right) \right]\; .
\ee
On the other hand, the fixed order ${\cal O}(\alpha_s)$
result for the partonic cross section for flavor $a$ is given
(at $Q_T\neq 0$) by~\cite{AEGM}:
\be
\label{qtb}
\frac{d\hat{\sigma}^{{\cal O}(\alpha_s)}}{dQ^2\,dQ_T^2} =
\sigma_a^{(0)} \,  \frac{\alpha_s(\mu)}{\pi} \,C_F
\,\left[ \frac{\ln(Q^2/Q_T^2)}{Q_T^2}
-\frac{3}{2Q_T^2}  + {\cal O}
\left( \ln(Q^2/Q_T^2) \right) \right] \; ,
\ee
where we have indicated the functional form of the first $Q_T$-suppressed
correction. It is easy to show that upon Fourier transformation of
the ${\cal O}(\alpha_s)$ in Eq.~(\ref{approb}) back to $Q_T$ space,
the first two terms $\ln^2 (\bar{b})
-\frac{3}{2}\ln (\bar{b})$ reproduce the first two
contributions to $d\hat{\sigma}/dQ^2\,dQ_T^2$ in Eq.~(\ref{qtb}).
The term $\propto \ln(\bar{b})/b$ in Eq.~(\ref{approb}), however,
yields a subleading contribution to the cross section, which is of the form
$\ln(Q_T^2/Q^2)/Q_T$, that is, down with respect to the leading and
next-to-leading logarithms, but more singular than the first
suppressed correction to the fixed order cross section in Eq.~(\ref{qtb})
which is just $\propto \ln(Q_T^2/Q^2)$. In other words, the
choice~(\ref{chiold}) introduces new dependence of the resummed cross
section on $Q_T$, not present in the cross section calculated at fixed order.
Even though this affects only subleading, integrable terms, which are beyond
the reach of our resummation anyway, this 
mismatch between Eqs.~(\ref{expb}) and (\ref{qtb}) at
${\cal O}(1/Q_T)$ produces a spurious logarithmic
singularity in $d\sigma/dQ^2\, dQ_T$, which is the
cross section we will compare to the data.  This problem is avoided to 
all orders by choosing $\chi$ as in Eq.\ (\ref{chinew}) with any $\eta>0$.

The function $\chi$ in Eq.~(\ref{chinew}) is only a slight modification 
of Eq.~(\ref{chiold}), but it has the property that at large $\bar{b}$ 
corrections to the leading term are suppressed as $1/\bar{b}^2$. 
In this way, no dependence of the form $\ln(Q_T^2/Q^2)/Q_T$ can arise in the
cross section in $Q_T$ space. Obviously, the general 
limits for $\bar{b}\to \infty$
and $\bar{N}\to\infty$ are the same as in~(\ref{chiold}).
It also turns out that this form of $\chi(\bar{N},\bar{b})$ leads
to analytic properties of the exponent in Eq.~(\ref{expdef})
that are consistent with the method described below for performing the inverse
transform in $b$.

We will analyze expansions of our final resummed cross section to one loop
in more detail in subsec.~\ref{oneloop}. Before doing so, we need to specify 
the coefficients ${\mathcal C}_{a/H}(Q,b,N,\mu,\mu_F )$ of Eq.~(\ref{crsec}).
\subsection{The coefficients ${\mathcal C}_{a/H}(Q,b,N,\mu,\mu_F )$}
The coefficients ${\mathcal C}_{a/H}(Q,b,N,\mu,\mu_F )$ in Eq.~(\ref{crsec})
are chosen to correspond to the jointly resummed cross section
in~\cite{lsvprl} for large $N$ and arbitrary $b$, and to 
$Q_T$ resummation for $b\to \infty$, $N$ fixed:
\be
\label{cpdf}
     {\mathcal C}_{a/H}(Q,b,N,\mu,\mu_F )
=   \sum_{j,k} C_{a/j}\left(N, \alpha_s(\mu) \right)\,
{\cal E}_{jk} \left(N,Q/\chi,\mu_F\right) \,
               f_{k/H}(N ,\mu_F) \; .
\ee
Here the $f_{j/H}(N ,\mu_F)$ are again the parton distribution 
functions for hadron $H$ at factorization scale $\mu_F$. In principle,
by analogy with standard $Q_T$ resummation~\cite{Coll85},
the scale for the strong coupling in the $C_{a/j}\left(N, \alpha\right)$
would be $Q/\chi$; however, at the NLL level we are considering here, it is 
legitimate to choose the ``large'' renormalization scale $\mu\sim Q$ and
to expand the $C_{a/j}\left(N, \alpha\right)$ to a finite order in 
$\alpha_s$. To first order, matching Eq.~(\ref{cpdf}) to the
large-$b$ behavior of the $Q_T$-resummed cross 
section~\cite{AEGM,Coll85,ds,CDG}, one has
\ba \label{caj}
C_{q/q}\left( N,\as \right) &=& 1+
\frac{\as}{4\pi} C_F \left(-8+\pi^2 +\frac{2}{N(N+1)} \right) \; =\;
C_{\bar{q}/\bar{q}}\left( N,\as \right)\; ,\\
\label{cajg}
C_{q/g}\left( N,\as\right) &=&\frac{\as}{2\pi}\frac{1}{(N+1)(N+2)}\; =\;
C_{\bar{q}/g}\left( N,\as\right)\;  .
\ea
Note that the coefficient $C_{q/g}$ is off-diagonal in flavor. It is indeed
a well-known feature in $Q_T$ resummation~\cite{Coll85,ds,AEGM,CDG}
that such non-diagonal terms also contribute to singular
behavior at $Q_T=0$. On the other hand, they do not incorporate
singularities at threshold, which is visible from the fact that
$C_{q/g}$ is suppressed at large $N$.

The evolution 
matrix ${\cal E} \left(N,Q/\chi,\mu_F\right)$ in Eq.~(\ref{cpdf}) 
results from the second term in Eq.~(\ref{Elog2}) that we chose
to absorb into the ${\cal C}$ coefficients. Compared to the large-$N$ 
limit used in that equation and relevant near threshold,
we can make an improvement here and replace the leading-$N$ 
part of the diagonal anomalous dimension in Eqs.~(\ref{leadN}) and
(\ref{Elog2}) by the {\em full}
anomalous dimension relevant for the scale evolution of 
parton densities, containing also all terms subleading in $N$:
\be
-2 A_a(\as) \ln\left( \bar{N}\right) - B_a(\as) \; \longrightarrow \;
\gamma_N (\as) \; .
\label{repl}
\ee 
At NLL level, where $A_a^{(1)}$, $B_a^{(1)}$, $A_a^{(2)}$
contribute, we need the first two terms in the perturbative
expansion of the anomalous dimension~\cite{andim}, $\gamma_N=\frac{\as}{\pi}
\gamma_N^{(0)}+\left(\frac{\as}{\pi}\right)^2 \gamma_N^{(1)}$.
Since the $\gamma_N^{(i)}$ are (in general) matrices, this procedure 
introduces terms that are parton non-diagonal, 
and thus leads to the matrix structure of ${\cal E}$ in terms
of an ordered exponential. As mentioned above, the interpretation
of ${\cal E}$ is simply the evolution of the parton densities 
from scale $\mu_F$ to scale $Q/\chi$, within the NLL 
approximation. In this way, it leads
to a resummation of collinear logarithms, some of which are
associated with partonic threshold, that is, are also proportional to 
$\ln \bar{N}$, while others are suppressed by ${\cal O}(1/N)$ or more, 
due to partonic mixing. Such a procedure is also familiar from standard $Q_T$ 
resummation~\cite{Coll85,CDG}, and the substitution in
Eq.~(\ref{repl}) thus provides a natural extension of our formalism 
for joint resummation away from threshold.
We may also interpret ${\cal C}_{a/H}(Q,b,N,\mu,\mu_F)$ as the Fourier-Mellin 
transform of a generalized parton distribution, at measured transverse 
momentum and energy fraction~\cite{lsv}. 
Eq.~(\ref{cpdf}) then has the interpretation of a
refactorization of  ${\cal C}_{a/H}$ into a coefficient function
$C_{a/j}$ and light cone parton distribution functions, 
$f_{j/H}$, at the scale $Q/\chi$.  

Explicit expressions for the solution of the standard evolution
equations for parton densities between scales $\mu_F$ and $Q/\chi$ 
can be found in~\cite{fp,bv} and can be used to 
determine the elements of the matrix ${\cal E}$. To achieve 
the exponentiation of the evolution terms -- despite the fact that
the matrices $\gamma_N^{(0)}$ and $\gamma_N^{(1)}$ do not
commute -- the iterative procedure derived in Ref.~\cite{bv} is 
particularly useful. The parameter that governs the evolution 
between scales $\mu_F$ and $Q/\chi$ is 
\be \label{sbeta}
\ln\left( \frac{\as(\mu_F)}{\as(Q/\chi)} \right)
=\ln (1-2 \beta) + \alpha_s(\mu) \left[
\frac{b_1}{b_0} \frac{\ln(1-2 \beta)}{1-2\beta}+
b_0 \ln \left( \frac{Q^2}{\mu^2} \right)
\frac{2 \beta}{1-2\beta} + b_0 \ln \left( \frac{Q^2}{\mu_F^2} \right)
\right] \; ,
\ee
where the right-hand-side is the expansion to NLL accuracy,
consistent with our approximation. It should be emphasized that in 
the above expression the scale $\mu_F$ appears through a single 
explicit logarithm that will serve to approximately 
cancel the $\mu_F$ dependence of the parton distributions in 
Eq.~(\ref{cpdf}), resulting in a decrease in $\mu_F$-dependence 
for the final resummed cross section in Eq.~(\ref{crsec}),
as compared to a fixed-order calculation. 

In all of these procedures, working in Mellin-$N$ moment space
is a great convenience, because it enables us to explicitly
express the evolution between the scales $\mu_F$ and $Q/\chi$
in terms of the parameter $\ln(\as(\mu_F)/\as(Q/\chi))$ in
Eq.~(\ref{sbeta}). In this way, we avoid 
the problem normally faced in $Q_T$ resummation that one needs to call 
the parton densities at scales far below their range of validity,
so that some sort of ``freezing'' (or related prescription) for handling
the parton distributions is required. As is evident from 
Eq.~(\ref{cpdf}), we only need the parton distribution functions
at the ``large'' scale $\mu_F \sim Q$, whereas normally in $Q_T$
resummation the product $\sum_k {\cal E}_{jk} \left(N,Q/\bar{b},\mu_F\right) \,
f_{k/H}(N ,\mu_F)$ is identified with $f_{k/H}(N ,Q/\bar{b})$.
An organization of the ${\cal C}$ coefficients in a form similar to ours
was first proposed in~\cite{CDG}. We finally note that the moment 
variable $N$ and, as will be discussed below, also the impact parameter 
$b$ are in general complex-valued in our approach, so that it is even more
desirable to separate the complex scale $Q/\chi$ from that in the
parton densities. In this way, it is not even necessary (albeit 
convenient) to have the parton densities in Mellin-$N$ moment
space, as provided in the code of~\cite{grv}. In fact, we can
generalize our analysis to any set of distribution functions, even if
specified only in $x$ space.  Details are discussed in Appendix~B.
\subsection{Finite-order ${\cal O}(\as)$ expansions of the resummed
cross section} 
\label{oneloop}
In this section, we compare expansions of our jointly resummed cross 
section to ``exact'' expressions for the electroweak annihilation 
cross section at $\CO(\as)$. We discuss first the limits 
$ N \rightarrow \infty,\ b=0$ and 
$ b \rightarrow \infty,\ N$ fixed separately. In these cases,
all results can be given analytically. 

The limit $ N \rightarrow \infty,\ b=0$ corresponds to 
pure threshold resummation for the total cross section. 
It is realized by integrating over $Q_T$ in Eq.~(\ref{crsec}), 
which sets $b=0$ there. Expansion of Eq.~(\ref{crsec}) to $\CO(\as)$,
using Eqs.~(\ref{expdef}), (\ref{caj}), (\ref{cajg}) and the
parton distribution functions evolved according to Eq.~(\ref{sbeta}),
then gives for the partonic cross sections in the $q\bar{q}$ and 
$qg$ scattering channels:
\begin{eqnarray}
\hat{\sigma}^{q\bar{q}}&=& \sigma_q^{(0)} \,\frac{\as}{2\pi}\,C_F\,\left\{
-4 \ln^2\bar{N}+6 \ln\bar{N}
-8+\pi^2+\frac{2}{N(N+1)} \right.
\nonumber \\
&& 
\hspace*{3cm}+\left.
\left[ \frac{2}{N(N+1)}+3-4S_1(N)\right] \;\left[
-2 \ln\bar{N}+
\ln\left(\frac{Q^2}{\mu_F^2} 
\right)\right]\right\}
\quad , \nonumber  \\
\hat{\sigma}^{qg}&=&  \sigma_q^{(0)} \,\frac{\as}{2\pi}\,
T_R\,\left\{ \frac{N^2+N+2}{N(N+1)(N+2)} \left[ -2 
\ln\bar{N}+\;\ln\left(\frac{Q^2}{\mu_F^2} \right) \right]+ 
\frac{2}{(N+1)(N+2)}\right\}\; ,
\end{eqnarray}
where $S_1(N)=\sum_{j=1}^N j^{-1}=\psi(N+1)+\gamma_E$, with $\psi$
the digamma function. Since we are interested in the near-threshold
region, we can expand this further to the large-$N$ limit. 
Using $\psi(N+1)= \ln N+1/(2N)+{\cal O}(1/N^2)$, we find:
\begin{eqnarray} \label{largen1}
\hat{\sigma}^{q\bar{q}}&=& \sigma_q^{(0)} \,\frac{\as}{2\pi}\,C_F\,\left\{
 4 \ln^2\bar{N}+
4\frac{\ln\bar{N}}{N}-8+\pi^2 +
\left[ 3-\frac{2}{N}-4\ln \bar{N}\right] \;\ln\left(\frac{Q^2}{\mu_F^2}
\right)\right\}+{\cal O}\left(\frac{\ln\bar{N}}{N^2} \right)\, , \nonumber
\\
\hat{\sigma}^{qg}&=&  \sigma_q^{(0)} \,\frac{\as}{2\pi}\,
T_R\; \frac{1}{N} \left[ -2 
\ln\bar{N}+\;\ln\left(\frac{Q^2}{\mu_F^2} \right) \right]
+{\cal O}\left(\frac{\ln\bar{N}}{N^2} \right)
\; .
\end{eqnarray}
The Mellin moments of the ``exact''
${\cal O}(\as)$ partonic cross sections can be found in~\cite{mssv}
and read:
\begin{eqnarray}
\hat{\sigma}^{q\bar{q}}_{\rm exact}&=& \sigma_q^{(0)} \,
\frac{\as}{2\pi}\,C_F\,\left\{
4 S_1^2(N) - \frac{4}{N(N+1)}S_1(N) 
+\frac{2}{N^2}+\frac{2}{(N+1)^2}-8 + \frac{4\pi^2}{3}  \right.
\nonumber \\
&& 
+\left.
\left[ \frac{2}{N(N+1)}+3-4S_1(N)\right] \;\ln\left(\frac{Q^2}{\mu_F^2}
\right)\right\}
\quad , \nonumber \\
\hat{\sigma}^{qg}_{\rm exact}&=&  \sigma_q^{(0)} \,\frac{\as}{2\pi}\,
T_R\,\left\{ -2 \frac{N^2+N+2}{N(N+1)(N+2)} S_1(N)
+\frac{N^4+11N^3+22N^2+14N+4}{N^2(N+1)^2(N+2)^2}  \right. 
\nonumber \\
&&+ \left.\;\frac{N^2+N+2}{N(N+1)(N+2)}\;\ln\left(\frac{Q^2}{\mu_F^2}
\right)\right\}\; .
\end{eqnarray}
At large $N$, this gives
\begin{eqnarray} \label{largen2}
\hat{\sigma}^{q\bar{q}}_{\rm exact}&=& 
\sigma_q^{(0)} \,\frac{\as}{2\pi}\,C_F\,\left\{
 4 \ln^2\bar{N}+
4\frac{\ln\bar{N}}{N}-8+\frac{4}{3}\pi^2 +
\left[ 3-\frac{2}{N}-4\ln \bar{N}\right] \;\ln\left(\frac{Q^2}{\mu_F^2}
\right)\right\}+{\cal O}\left(\frac{\ln\bar{N}}{N^2} \right)\, , \nonumber \\
\hat{\sigma}^{qg}_{\rm exact}&=&  \sigma_q^{(0)} \,\frac{\as}{2\pi}\,
T_R\; \frac{1}{N} \left[ -2 
\ln\bar{N}+\;\ln\left(\frac{Q^2}{\mu_F^2} \right) \right]
+{\cal O}\left(\frac{\ln\bar{N}}{N^2} \right)\; .
\end{eqnarray}
Comparing Eqs.~(\ref{largen1}) and (\ref{largen2}), we see that 
at large $N$ the expansion of the resummed cross section correctly 
reproduces the ${\cal O}(\as)$ result, including even all terms that
are down by $1/N$. The only difference between (\ref{largen1}) and 
(\ref{largen2}) is in the term $\propto \pi^2$. This difference is
due to our choice of the coefficient $C_{q/q}$ in Eq.~(\ref{caj}),
for which we employed a form that is more standard in $Q_T$, rather than 
in threshold,
resummation.  A closer inspection of our resummed eikonal exponent,
Eq.~(\ref{eikexpdef}), reveals that its Bessel functions result in
different contributions $\propto \pi^2$ in the two limits
$N\to \infty$ and $b\to \infty$, just as needed to explain
the deficiency between Eqs.~(\ref{largen1}) and (\ref{largen2}).
In other words, we could modify our expansion of the exponent 
into logarithms somewhat (by suitably redefining $\chi$), so 
that Eq.~(\ref{largen1}) would automatically have the correct
coefficient of $\pi^2$. On the other hand, the terms associated
with $\pi^2$ are beyond NLL, which is the scope
of the present anlysis, and we therefore do not implement this change
here. 

It is worth pointing out that the reason why we correctly reproduce
all terms suppressed as $1/N$ in Eqs.~(\ref{largen1}) and (\ref{largen2})
is our treatment of evolution in Eq.~(\ref{repl}). As was discussed 
in~\cite{CDG1}, the leading $\ln(\bar{N})/N$ terms are associated  
with collinear non-soft emission; it is therefore natural that
they can be generated from evolution of the parton densities between 
the scales $\mu_F$ and $Q/\bar{N}$, as embodied in Eq.~(\ref{sbeta}) at
$b=0$. In this way, our joint resummation corectly includes the
leading $\as^k \ln^{2k-1}\bar{N}/N$ terms to all orders. Because threshold 
resummation has a relatively modest effect for vector boson production 
in the kinematic region explored at the Tevatron, we leave for future 
work a more complete comparison of our resummed expression to fixed order, 
beyond these ${\cal O}(\as)$ considerations.

In the limit $ b \rightarrow \infty,\ N$ fixed, our formulas 
smoothly turn into those for standard $Q_T$ resummation. For
the one-loop expansion of the jointly resummed cross section
we find
\begin{eqnarray}
\hat{\sigma}^{q\bar{q}}&=& \sigma_q^{(0)} \,\frac{\as}{2\pi}\,C_F\,\left\{
-4 \ln^2 \bar{b} +6 \ln\bar{b}-8+\pi^2+\frac{2}{N(N+1)}
\right. \\
&& 
\hspace*{2cm} + \left.
\left[ \frac{2}{N(N+1)}+3-4S_1(N)\right] \;\left[ -2 \ln\bar{b}+
\ln\left(\frac{Q^2}{\mu_F^2}
\right)\right]\right\}+{\cal O}(\ln\bar{b}/b^2)
\quad , \nonumber  \\
\hat{\sigma}^{qg}&=&  \sigma_q^{(0)} \,\frac{\as}{2\pi}\,
T_R\,\left\{ \frac{N^2+N+2}{N(N+1)(N+2)} \left[ -2 
\ln\bar{b}+\;\ln\left(\frac{Q^2}{\mu_F^2} \right) \right]+ 
\frac{2}{(N+1)(N+2)}\right\}+{\cal O}(\ln\bar{b}/b^2)\; ,\nonumber
\end{eqnarray}
in full agreement with the expressions for the large-$b$ limit of the 
``exact'' ${\cal O} (\as)$ result derived in~\cite{AEGM}.

\begin{figure}[h]
\begin{center}
\epsfig{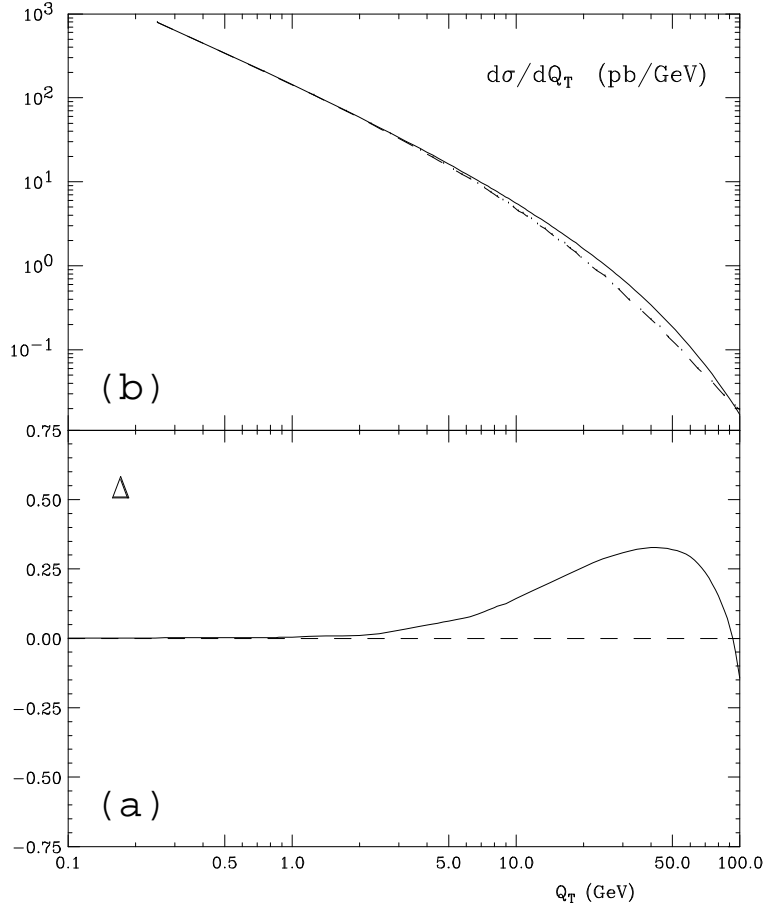}
\caption{(a) Fractional deviation $\Delta$ (as defined in Eq.~(\ref{frdev}))
between the ``exact'' $\CO (\as)$ result and the $\CO (\as)$ 
expansion of the jointly resummed cross section. We consider here
Z boson production at the Tevatron; the cross section has been integrated
over $66<Q<116$ GeV. 
(b) Comparison of $d \sigma^{\rm fixed(1)} /d Q_T$ (solid)
and $d\sigma^{\rm exp(1)}/d Q_T$ (dashed) on absolute scale.
\label{resfixcomp}
}
\end{center}
\end{figure}

We do not present closed expressions for arbitrary large $N$ and $b$, but
we can easily compare numerically the exact $\CO (\as)$ result with the 
expansion of Eq.~(\ref{crsec}) to $\CO(\as)$. Fig.~\ref{resfixcomp}(a) 
shows the fractional deviation 
\begin{equation}
\Delta\equiv\left[ 
{d \sigma^{\rm fixed(1)} \over d Q_T} -
{d\sigma^{\rm exp(1)} \over d Q_T}\right] \;/\; {
d \sigma^{\rm fixed(1)} \over d Q_T} \,,
\label{frdev}
\end{equation}
where 
$d \sigma^{\rm fixed(1)} /d Q_T$ is the ``exact'' 
${\cal O}(\as)$ cross section and  $d\sigma^{\rm exp(1)}/d Q_T$ 
denotes the one-loop  expansion of the resummed expression.
Fig.~\ref{resfixcomp}(b) compares $d \sigma^{\rm fixed(1)} /d Q_T$
and $d\sigma^{\rm exp(1)}/d Q_T$ individually.
Note the excellent agreement in the region $Q_T<10$ GeV where 
resummation is necessary. Beyond $10$ GeV, the agreement is naturally 
less exact but still good. This is the region where matching
to finite order is appropriate, to which we will turn now.
\vspace*{-2mm}
\section{Inverse transforms and matching}
In this section we take the remaining steps
necessary to apply the joint resummation formalism 
phenomenologically. This includes specifying
a prescription for performing the inverse integrals in~(\ref{crsec}),
as well as implementing a procedure for matching resummed and finite-order
results. In addition, we will consider \nprt effects resulting
from the strong coupling at small momentum scales.

\subsection{Inverse transforms}
\label{sec:invtranf}

When performing the inverse $N$ and $b$ transforms, special attention
has to be paid to the singularity in the resummed exponent,
Eqs.~(\ref{expdef})--(\ref{hsubadef}), at $\beta = 1/2$. For  $\eta=1/4$, the 
singularity occurs for
\be
\chi(\bar{N},\bar{b})=\bar{b} + \frac{\bar{N}}{1+\frac{\bar{b}}{4 \bar{N}}}
=\exp\left[1/(2b_0\alpha_s(\mu))\right]
\equiv \rho_L\; ,
\ee
and is a manifestation of the Landau pole in the strong coupling.
The exponent is also ill-defined when 
$\chi=0$ and
infinity, i.e., at $\bar{b} =-2 \bar{N}$ and $ \bar{b} =-4 \bar{N}$,
respectively.  We note that the choice $\eta=1/4$ is made
simply to reduce $\chi=0$ to a linear
relation between $\bar{b}$ and $\bar{N}$.  This is not
an essential simplification, but it makes the following
analysis slightly more convenient.  In the following we
sketch the application of the method of  \cite{lsvprl} to this jointly 
resummed cross section. As far as the Landau pole
is concerned, it is simultaneously in the spirit of both
the ``minimal'' prescription proposed for pure threshold
resummation in~\cite{cmnt},
and the ``principal value resummation", described in \cite{cspv}.

\begin{figure}[h]
\vspace*{2mm}
\begin{center}
\epsfig{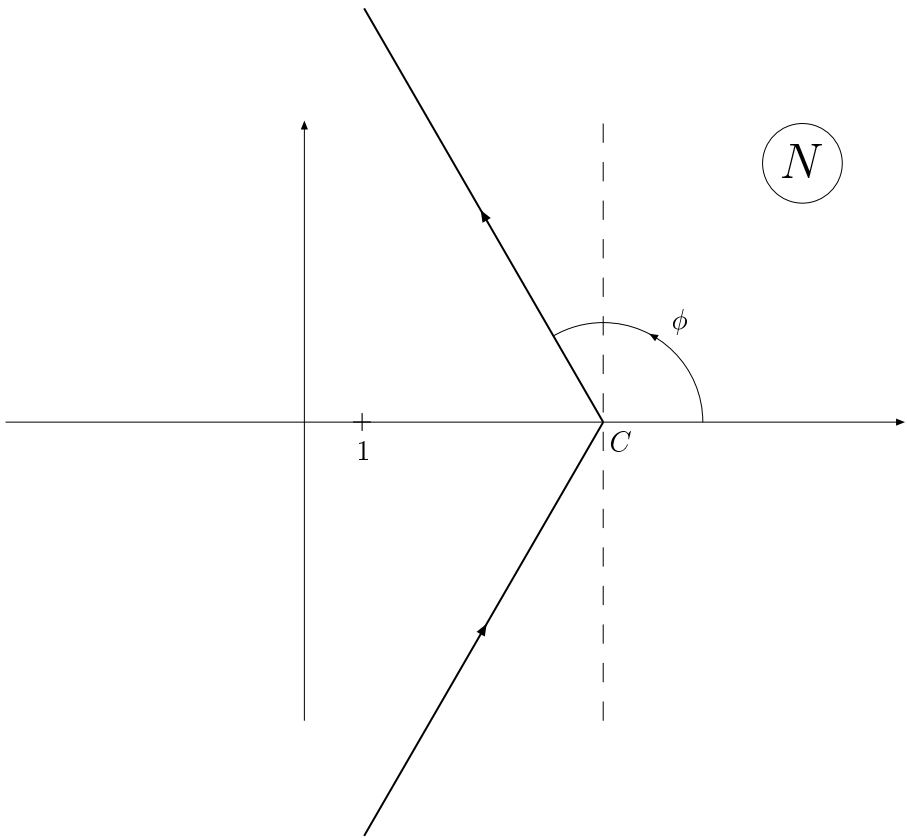}
\caption{Choice of contour for Mellin inversion.
\label{Ncont}}
\end{center}
\end{figure}

The contour for the inverse Mellin transform is chosen to be bent
at an angle $\phi$ with respect to the real axis and is parameterized
as follows (see Fig.~\ref{Ncont}):
\be
\label{cont}
N = C+ z {\rm e}^{\pm i \phi} \; ,
\ee
where the upper (lower) sign applies to the upper (lower)
branch of the contour, with $0\leq z\leq \infty$ ($\infty\geq z\geq 0$).
For $\phi>\pi/2$, this results in an exponentially convergent integral
over $N$ in the inverse transform, Eq.\ (\ref{crsec})
for all $\tau<1$ \cite{cs,cmnt}.
For the moment, we do not specify the constant $C$, except that
it has to lie to the right of the rightmost singularity of the parton
distribution functions.  At any finite order in perturbation theory,
all values of $C>0$, and all $\pi>\phi>\pi/2$ are equivalent.  In the
resummed cross section, however, the singularity at $\chi=\rho_L$ introduces
a power-suppressed ambiguity in the transform,
which we resolve by choosing $C<\rho_L$ \cite{cmnt} .

As mentioned above, the position of the Landau pole depends on
both $N$ and $b$. The  $d^2b$ integral in Eq.~(\ref{crsec}) can be
written as
\be
\label{bint}
\int d^2 b \;e^{i{\vec{Q}_T}\cdot {\vec{b}}}\,f(b)\;=\;  2 \pi\,
\int_0^{\infty} \, db\,b\, \,J_0(bQ_T) \,f(b) \; ,
\ee
where $J_0$ is the Bessel function. Hence, as it stands, the
integration over $b$ lies on the positive real axis. This would
imply that one would never be able to avoid hitting the Landau
pole, since for any choice of the parameter $C$ in Eq.~(\ref{cont})
there is a $b>0$ for which $\chi(\bar{N},\bar{b})=\rho_L$.
This problem is of course well-known from standard $Q_T$ resummation. The 
procedure usually adopted is to prevent $b$ from becoming too large by
evaluating the resummed cross section at $b_{\ast}=b/
\sqrt{1+b^2/b_{\rm max}^2}$~\cite{Coll85,Coll82,ds,AK,ly,Res}, 
at the expense of introducing
a new parameter $ b_{\rm max}$. To avoid introducing a new 
parameter, we treat the $b$ integral in a manner analogous to the $N$
integral above, avoiding the Landau singularity
on a contour that produces an exponentially convergent
integral for all $Q_T>0$.

\begin{figure}[h]
\vspace*{2mm}
\begin{center}
\epsfig{file=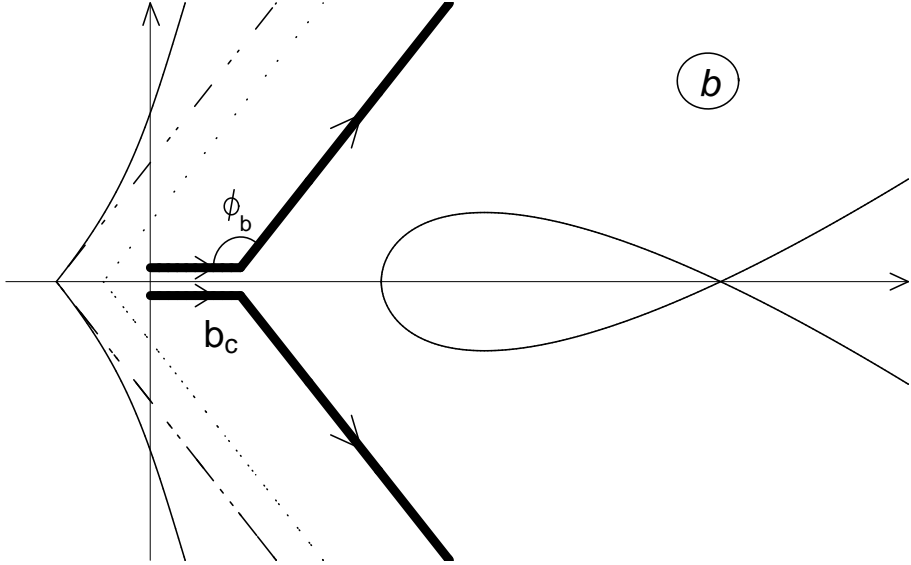,width=12cm}
\caption{Choice of contour for $b$ integration (thick solid lines).
The straight sections of the contour from $0$ to $b_c$ are to 
be interpreted as on the positive real axis. The remaining curves represent
lines of singularity discussed in the text.}
\label{bcont}
\end{center}
\end{figure}

Were the Landau pole not present we could, instead of performing
the $b$ integral along the real axis, use Cauchy's theorem and
divert it into complex $b$ space along either of the two solid lines
in Fig.~\ref{bcont}, under the condition that the integrand falls off
sufficiently fast at large $|b|$ and that there be no
contribution to the integral at infinitely large real part. To
achieve this, we have to split up Eq.~(\ref{bint}) as~\cite{lsvprl}
\be
2 \pi\,
\int_0^{\infty} \, db\,b\, \,J_0(bQ_T) \,f(b)    =
\pi\, \int_0^\infty db\, b\, \left[\, h_1(bQ_T,v) + h_2(bQ_T,v)\,
\right]\,f(b) \, ,
\label{J0split}
\ee
where we introduce two auxiliary functions $h_{1,2}(z,v)$, related
to Hankel functions and defined in terms of an arbitrary real,
positive parameter $v$ by integrals in the complex 
$\theta$-plane~\cite{grad}:
\ba
h_1(z,v)
&\equiv& - {1\over\pi}\
\int_{-iv\pi}^{-\pi+iv\pi}\, d\theta\, {\rm e}^{-iz\, \sin\theta}\; ,
\nonumber\\
h_2(z,v)
&\equiv& - {1\over\pi}\
\int^{-iv\pi}_{\pi+iv\pi}\, d\theta\, {\rm e}^{-iz\, \sin\theta}\, .
\label{hdefs}
\ea
For $h_1$, we parameterize $\theta =
-i v \pi + x_{\theta} \pi (-1+2 i v)$ $(0\leq x_{\theta} \leq 1)$,
while for $h_2$, $\theta = -i v \pi + x_{\theta} \pi (1+2 i v)$
$(1\geq x_{\theta}\geq 0)$. 
The $h_{1,2}$ become the usual Hankel functions $H_{1,2}(z)$ in the
limit $v\rightarrow \infty$.  We note that this convergence to
the Hankel functions is extremely rapid, since the dependence
on the variable $v$ is suppressed by the exponential of an 
exponential for all finite $z$. 
The $h_{1,2}$ are finite for any finite
values of $z$ and $v$.  Their sum is always: $h_1(z,v)+h_2(z,v)=2J_0(z)$,
independent of $v$. The utility of the $h$-functions is that they distinguish
positive and negative phases in Eq.\ (\ref{J0split}), making it possible
to treat the $b$ integral as the sum of the two contours
in Fig.~\ref{bcont}, the one associated with $h_1$ ($h_2$)
corresponding to closing the contour in the upper (lower) half plane.

The virtue of this technique for the $b$ integration is that we can
choose the contours to avoid the Landau pole. We simply need to make
sure that the $b$ contour never intersects the trajectories defined by
$\chi(\bar{N},\bar{b}) = \rho_L\;$,
shown by the two light solid curves in Fig.~\ref{bcont}.
As mentioned earlier, singularities also arise for $\bar{b}=-2\bar{N}$ and
$\bar{b}=-4\bar{N}$. These contours
are shown in Fig.~\ref{bcont} by the dotted line and the dash-dotted line,
respectively.  Parameterizing the upper $b$ contour as
\be
\label{upperb}
C_1: \quad b =\left\{ \begin{array}{ll}
   t & (0\leq t\leq b_c) \\
b_c - t  {\rm e}^{-i \phi_b} & (0\leq t\leq \infty)
\end{array} \right.
\ee
and the lower one as
\be
\label{lowerb}
C_2: \quad b =\left\{ \begin{array}{ll}
   t & (0\leq t\leq b_c) \\
   b_c - t  {\rm e}^{i \phi_b} & (0\leq t\leq \infty)
\end{array} \right. \; ,
\ee
we  choose the parameters $b_c$ and $\phi_b$
such that none of the branches intersects any of the ``forbidden''
lines in Fig.~\ref{bcont}. A typical choice is also shown
in Fig.~\ref{bcont} by the thick solid lines. The parameters $C$ in
Eq.~(\ref{cont}) and $b_c$ in Eqs.~(\ref{upperb}), (\ref{lowerb})
are arbitrary as long as  $0<\left( {C \over 1+ b_c Q/8C}  +
{b_c Q \over 2}\right){\rm e^{\gamma_E}} < \rho_L$.
In this way, our full expression for the cross section
in terms of inverse transforms, Eq.\ (\ref{crsec}),
becomes
\ba
\label{crsec2}
     \frac{d\sigma_{AB}^{\rm res}}{dQ^2\,dQ_T^2}
     &=&    \sum_a \sigma_{a}^{(0)}\,
\int_{C_N}\, \frac{dN}{2\pi i} \,\tau^{-N}\;  \Bigg\{\; 
\int_{C_1} \frac{db\, b}{4\pi} \,
h_1(bQ_T,v)\, \bar W_{AB}^{(a)}(Q,b,N,\mu,\mu_F)\,    \nonumber \\
&\ & \hspace{5mm} + 
\int_{C_2} \frac{db\, b}{4\pi} \,
h_2(bQ_T,v)\, \bar W_{AB}^{(a)}(Q,b,N,\mu,\mu_F) 
\; \Bigg\}\; ,
\ea
with  
\ba
\label{Wdef}
\bar W_{AB}^{(a)}(Q,b,N,\mu,\mu_F) 
&=& \exp\left[ \, E_{a\bar a}^{\rm PT} (N,b,Q,\mu)\,\right] 
\nonumber\\
&\ & \quad\quad \times \, 
\sum_{j,k} C_{a/j}\left(N, \alpha_s(\mu) \right)\,
{\cal E}_{jk} \left(N,Q/\chi,\mu_F\right) \,
               f_{k/A}(N ,\mu_F)
\nonumber\\
&\ & \quad\quad \times \, 
\sum_{\bar{j}\bar{k}} C_{\bar{a}/\bar{j}}\left(N, \alpha_s(\mu) \right)\,
{\cal E}_{\bar{j}\bar{k}} \left(N,Q/\chi,\mu_F\right) \,
               f_{\bar{k}/B}(N ,\mu_F)\, ,
\ea
for initial hadrons $A$ and $B$.

This choice of contours in complex transform space is
completely equivalent to the original form, Eq.\ (\ref{bint}) when
the exponent is evaluated to finite order in perturbation theory.
It is a natural extension of
the $N$-space contour redefinition above \cite{cs,cmnt}, using a 
generalized ``minimal'' \cite{cmnt} exponent, Eq.\ (\ref{expdef}).
As we stressed earlier, joint resummation with its contour
integration method  provides an  alternative to
the standard $b$ space resummation.  Joint resummation has built-in
\prt treatment of large $b$ values, eliminating the need for a
$b_*$ or other prescription for the exponent, or for a freezing of the
scale of parton distributions at large $b$ or low $Q_T$.

\begin{figure}[t]
\begin{center}
\mbox{\epsfig{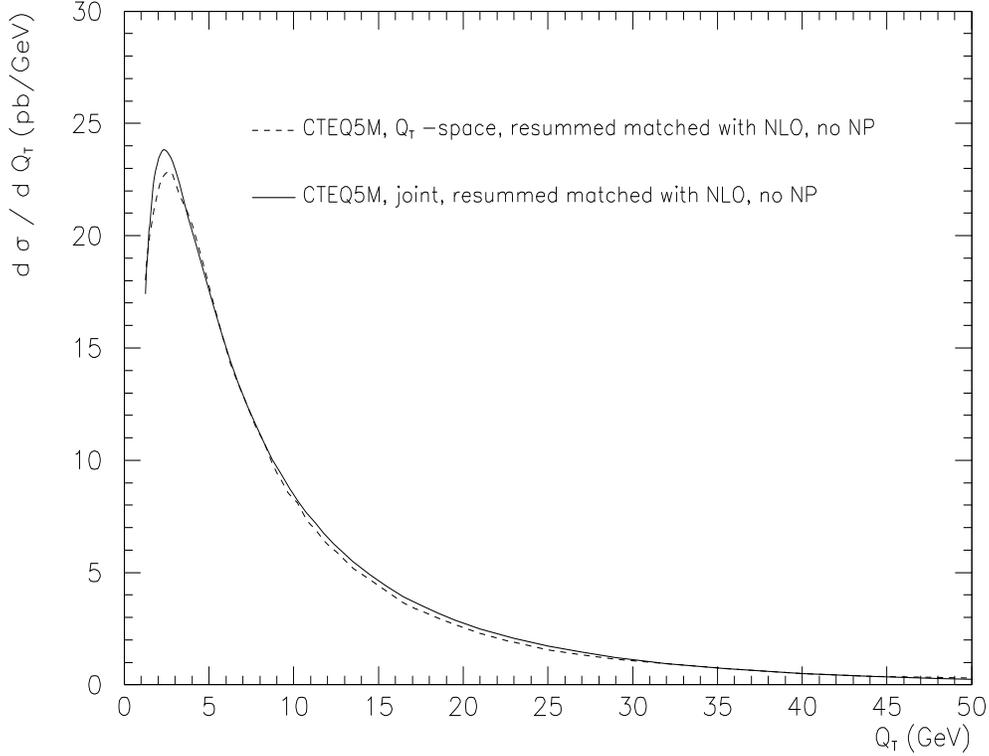}}
\end{center}
\caption{$Q_T$ distribution for Z production at $\sqrt s=1.8$ TeV calculated
using the narrow width approximation ($Q=M_Z$). The $Q_T$
    space method result and the joint resummation method result are matched to
    the cross section at ${\cal O}(\alpha_s)$. 
The CTEQ5M~\cite{cteq} parton distributions have
been used.
\label{matching}}
\end{figure}

To examine the relevance of the large-$b$ contributions,
we now compare the jointly resummed $Q_T$ distribution
matched to the ${\cal O}(\alpha_s)$ perturbative result~\cite{AEGM}
and the $Q_T$ space resummed~\cite{KS} distribution also matched to the 
${\cal O}(\alpha_s)$
distribution (see the next section for details of the
matching procedure). The $Q_T$ space resummation formalism originates
from $b$ space resummation and can be viewed as a very good approximation
of the latter. Unlike the standard $b$ space technique,
direct $Q_T$ resummation, like the contour method just described, yields
a result even without any \nprt input for nonzero values of
$Q_T$.  The two approaches are compared in Fig.~\ref{matching}
for the case of Z boson production at Tevatron energy.  
As expected, the two distributions differ mostly
at the very low-$Q_T$ end of the spectrum.

\subsection{Matching to finite order}
\label{sec:match}

A resummed $Q_T$ distribution provides theoretical predictions for the
small $Q_T$ region; in the large $Q_T$ regime one mainly relies on
fixed-order perturbation theory. Thus a description
of the intermediate $Q_T$ region requires a consistent matching between
the two results that avoids double counting.
For the joint resummation we adopt a matching prescription
first proposed in standard $b$ space resummation~\cite{AK}:
\begin{equation}
{d \sigma \over d Q^2 d Q_T^2} = {d \sigma^{\rm res} \over d Q^2 d Q_T^2}
-  {d\sigma^{\rm exp(k)} \over d Q^2 d Q_T^2} +
{d \sigma^{\rm fixed(k)} \over d Q^2 d Q_T^2} \,,
\label{joint:match}
\end{equation}
where $d \sigma^{\rm res}/d Q^2 d Q_T^2$ is given in Eq.~(\ref{crsec}) and,
as before, 
$d\sigma^{\rm exp(k)}/d Q^2 d Q_T^2$ denotes the terms resulting from
the expansion of the resummed expression in powers of
$\as(\mu)$ up to the order $k$ at which the fixed-order cross section
$d \sigma^{\rm fixed(k)} /d Q^2 d Q_T^2$  is taken (in practice,
$k=1$ (see~\cite{AEGM}) or $k=2$ (see~\cite{NNLO})).

Alternatively, one can formulate the matching procedure in the following
way~\cite{Coll85,AEGM}:
\begin{equation}
{d \sigma \over d Q^2 d Q_T^2} = {d \sigma^{\rm res} \over d Q^2 d Q_T^2} +
Y_{\rm finite} \; ,
\label{qt:match}
\end{equation}
$Y_{\rm finite}$ standing for the finite part of the fixed-order
distribution, i.e. the part remaining after the singular behavior
$\propto  \as^k \ln^m (Q^2 /Q_T^2)/Q_T^2,\;  (2k -1 \geq m \geq 0)$
at $k$th order of perturbation theory (in our case, $k=1,2$)
has been taken out of the fixed-order cross section. For pure
$Q_T$ resummation, there is no difference between the two ways of
matching in Eqs.~(\ref{joint:match}), (\ref{qt:match}), as long as
$d\sigma^{\rm exp(k)}/d Q^2 d Q_T^2$ coincides identically with the
singular part of the fixed-order cross section. For our joint resummation,
however, $d\sigma^{\rm exp(k)}/d Q^2 d Q_T^2$  also contains terms
that are non-singular in $Q_T$ (as in Eq.\ (\ref{approb})), which implies 
that the matching is most naturally performed
according to Eq.~(\ref{joint:match}) in conjugate ($b,N$) space, in
order to avoid double counting.

Notably, joint resummation with the choice of $\chi(\bar{N},\bar{b})$
in Eq.~(\ref{chinew}) and the matching prescription~(\ref{joint:match})
returns a positive cross section even in the high $Q_T$ regime.
This is in contrast to pure $Q_T$ resummation, which notoriously yields a
negative answer at large $Q_T$. To avoid the latter feature,
usually an additional switch between the matched cross section
and the fixed-order result is implemented in pure $Q_T$ resummation.
A nice feature of joint resummation is that this is
not necessary here; Eq.~(\ref{joint:match}) is all we need all the way to
$Q_T$ of order $Q$.  (Of course, for $Q_T\gg Q$ a further resummation
is necessary \cite{Ber01}.)
The improved large-$Q_T$ behavior of the matched cross section
can be traced back to the behavior of the jointly resummed part
$d \sigma^{\rm res}/d Q^2 d Q_T^2$ itself, which also
remains positive for all values of  $Q_T$.  This in turn results
from the behavior of the function $\chi(\bar N,\bar b)$ at
the small values of $b$ relevant to high $Q_T$.  

A simple argument shows how the small-$b$ behavior
of $\chi$ ensures that the cross section 
remains positive at large $Q_T$ for
the jointly resummed cross section, and also why it goes negative
in pure $Q_T$ resummation in impact parameter space.  
Consider first pure $Q_T$ resummation, 
which is equivalent to the choice $\chi=\bar b$.
For any $\chi$, the one-loop expansion
of the resummed exponent in impact parameter space is given by 
Eq.\ (\ref{expb}).   When $\chi =\bar b$ simply
($\bar N=0$ in Eq.\ (\ref{approb})), this expression is just as singular 
at $b=0$ as for $b\rightarrow \infty$ \cite{Fri98}.  This produces a 
spurious  suppression
for $b$ small compared to $1/Q$, which is present as well
in the full exponent.  Because the $b$ integral is dominated by $b\sim 1/Q_T$,
this effect is unimportant when $Q_T$ is small.  On the other hand, when
$Q_T$ is large, of the same order as $Q$, the suppression of
small $b$ removes an important contribution from the $b$ integral.
Along the deformed contours of Fig.\ \ref{bcont}, the situation is
particularly simple.  Referring to Eq.\ (\ref{crsec2}), the sum of the
functions $h_1(bQ_T,v)$ and $h_2(bQ_T,v)$ starts  out at
2 for $b=0$, but
both begin to  oscillate and to decrease exponentially, when $b\sim 1/Q_T$.
A spurious suppression times $\bar W_{AB}^{(a)}(Q,b,N,\mu,\mu_F)$
for $b\le 1/Q\sim 1/Q_T$ will thus eliminate positive contributions
to the transform, leaving over negative contributions, in the
limited range of $b$ in which $h_1$ and $h_2$ are themselves of order unity.  
If, on the other hand, we pick $\chi$ as in Eq.\ (\ref{chinew}),
then we ensure that the spurious
suppression at small $b$ characteristic of pure $b$-space resummation is 
absent.  The result is a positive cross section for large $Q_T$. 

\subsection{Nonperturbative input}

In most applications of $b$ space resummation,
the \prt component is supplemented by a $Q$-dependent Gaussian in
impact parameter space, to tune the overall influence
of nonperturbative dynamics \cite{Coll85,ds,AK,ly,Res}. 
Joint resummation implies similar effects in the $Q_T$ distribution at 
small $Q_T$.   The starting point is the full
NLL exponent~\cite{lsv} given in Eq.~(\ref{eikexpdef}).
We expand the Bessel functions in Eq.~(\ref{eikexpdef}) to derive
an explicit form for the leading $b$- and $N$-dependence of the
resummed exponent, times an integral of the
anomalous dimension $A_a$ over soft transverse momenta,
defined by a cutoff $\lambda > \Lambda_{\mathrm{QCD}}$,
\be
E_{a\bar a}^{\rm eik} (N,b,Q,\mu,\mu_F,\lambda)
\sim \left( -{b^2 \over 2} + {2 N^2 \over Q^2} \right)
\int_0^{\lambda^2}
d k_T^2 A_a(\as(k_T)) \ln \left( {Q \over \bar N k_T} \right)  \, .
\label{explowkt}
\ee
In this way we derive a standard Gaussian form in $b$
for a multiplicative, \nprt smearing function in $b$ space,
to account for effects from small $k_T$.  The new feature here
is that the coefficient of $b^2$ is essentially identical to that of
the threshold-related power correction $(N/Q)^2$.
In our numerical applications
below, we will be rather far away from threshold, and as a result we will
be mainly sensitive to moderately small $N$. Hence we will retain only
the term $\sim b^2$ in Eq.~(\ref{explowkt}). Thus our \nprt smearing function
reverts to the purely Gaussian one usually used in $Q_T$ resummation, and
we will make the following replacement in the resummed cross section:
\be
E_{a\bar a}^{\rm PT} (N,b,Q,\mu) \longrightarrow
E_{a\bar a}^{\rm PT} (N,b,Q,\mu) - g b^2 \; ,
\label{replace}
\ee
where $g$ is a parameter to be determined by comparison to data. Note
that Eq.~(\ref{explowkt}) implies that, as usual, $g$ has a component
that depends logarithmically on $Q$ \cite{Coll85,ds,AK,ly,Res,Taf01}.

\section{Vector boson production in the framework \\
of joint resummation}

With the developments discussed above, the joint resummation formalism
becomes a practical tool for the description of electroweak annihilation.
Here we consider Z boson production at the Tevatron collider. Recent
data on the $Q_T$ distribution of the produced Z bosons are available from
both CDF~\cite{cdf} and D0~\cite{d0} experiments and have reached
a good level of precision. The overall normalization of the data will be
treated as a free parameter in our analysis, and will be varied
within the quoted experimental errors.

All results are obtained using CTEQ5M parton distribution
functions~\cite{cteq}, in the manner described in Appendix~\ref{app:xpdfs}.
We choose the factorization and renormalization scales
$\mu=\mu_F=Q$. The numerical values of the electroweak parameters we use are
as follows: $M_Z=91.187$ \GeV, $\sin^2 \theta_W= 0.224$, 
$\Gamma_Z=2.49$ \GeV. Note that the experimental data sets have been
integrated over finite regions in $Q$, $66<Q<116$ GeV for CDF and
$75<Q<105$ GeV for D0. The inverse transforms are performed as described in
Section~\ref{sec:invtranf}, with the following choice of the contour
parameters in Figs.\ 1 and 2: $\phi=\phi_b=25/32 \;\pi $, $C=1.3 $, 
$b_c=0.2/Q$.  Of course we
are free to choose these parameters differently, as long
as they are such that the structure of the contours
as depicted in Fig.~\ref{bcont} is maintained.
Finally, as mentioned above, we choose $\eta=1/4$ in Eq.\ (\ref{chinew}).
We have checked that the result is quite insensitive to this 
choice, at the order of a percent when $\eta$ is
changed by a factor of 2 at $Q_T=$ 4 GeV.  

\begin{figure}[h]
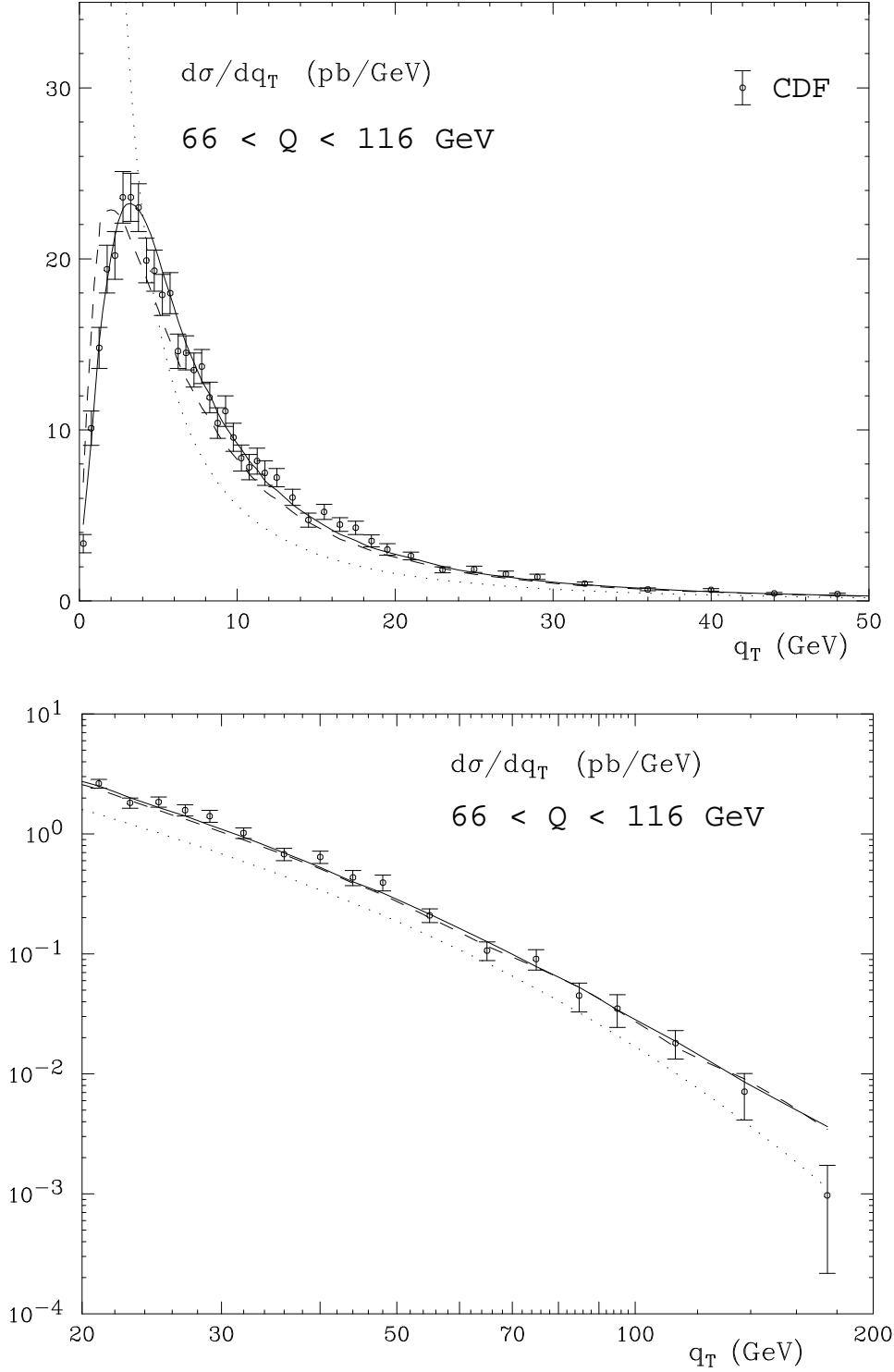

\begin{center}

\epsfig{figure=cdf.epsi,angle=90,width=12cm}

\vskip 5mm

\hspace*{-5mm}
\epsfig{figure=cdf_largeqt.epsi,angle=90,width=12.7cm}

\end{center}

\vspace*{-4mm}
\caption{CDF data {\protect \cite{cdf}} on Z production compared to joint 
resummation predictions (matched to the ${\cal O}
(\as)$ result according to Eq.~(\ref{joint:match}))
without \nprt smearing (dashed) and with Gaussian smearing using the
nonperturbative  parameter $g=0.8$ GeV$^2$ (solid). The normalizations
of the curves have been adjusted in order to give an optimal
description; see text.  The dotted and dash-dotted lines show the 
fixed-order results at ${\cal O}(\alpha_s)$ and ${\cal O}(\alpha_s^2)$,
respectively. The lower plot makes the large $Q_T$ region more visible.}
\label{fig:cdf}
\end{figure}

\begin{figure}[h]
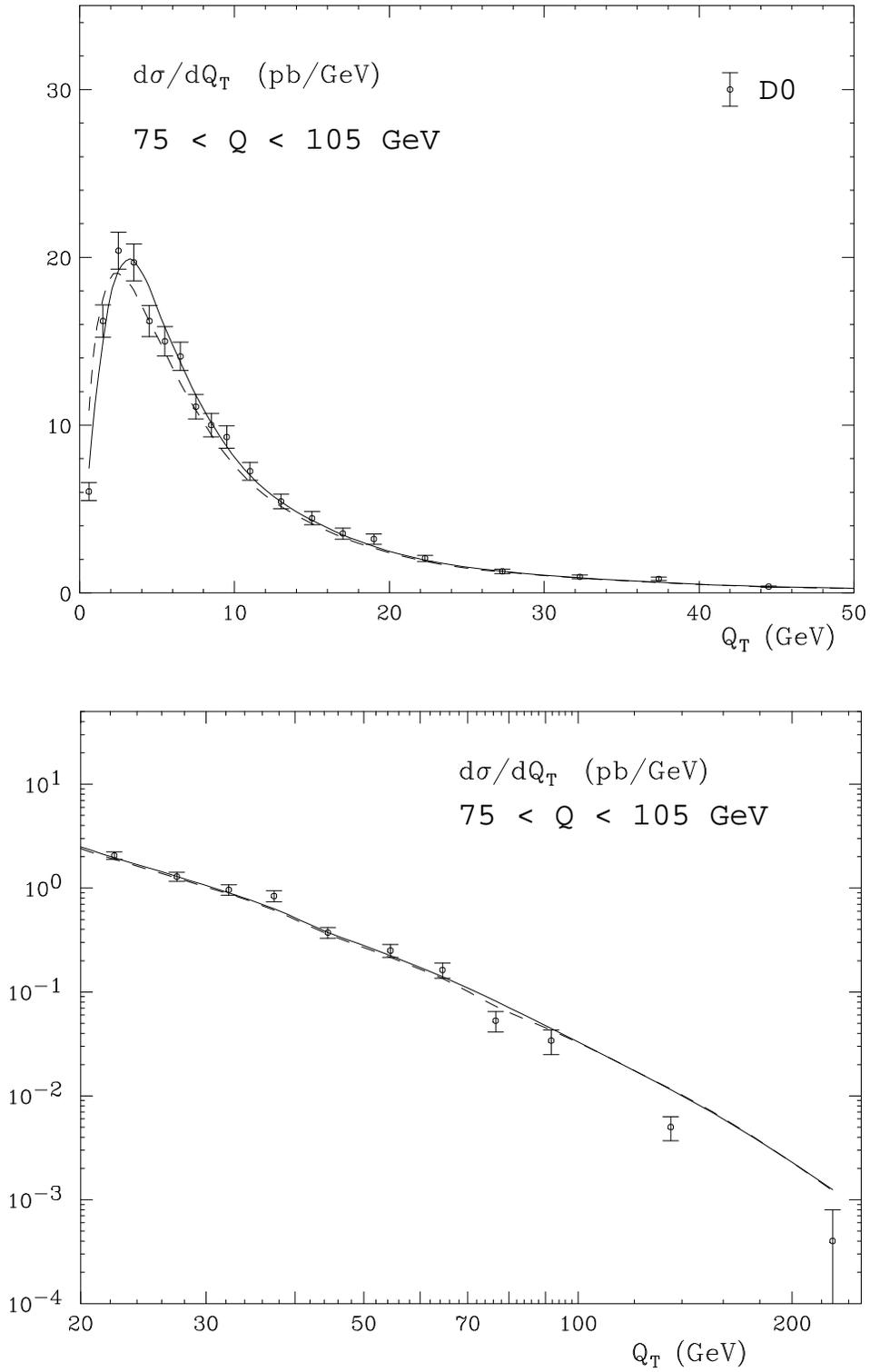

\begin{center}

\epsfig{figure=d0.epsi,angle=90,width=12cm}

\vskip 8mm

\hspace*{-7.6mm}
\epsfig{figure=d0_largeqt.epsi,angle=90,width=12.5cm}

\end{center}
\caption{Same as Fig.~\ref{fig:cdf}, but compared to the D0
data {\protect \cite{d0}}.}
\label{fig:d0}
\end{figure}

As we pointed out before, an attractive feature of the joint resummation 
with transforms defined as above is 
that we can obtain predictions that have no dependence
on any additional nonperturbative parameter.  Results of this form are shown
by the dashed lines in Figs.~\ref{fig:cdf} and~\ref{fig:d0}. In both
cases, we have adjusted the normalization of the theory curve so
that the $\chi^2$ of the comparison between data and theory becomes
minimal. For the CDF data, this normalization factor is 1.035, for D0
it is 0.96. It is evident from Figs.~\ref{fig:cdf} and~\ref{fig:d0}
that our ``purely perturbative'' predictions correctly reproduce
the trend of the data over most of the measured region in $Q_T$, but
peak at too small $Q_T$. This is no surprise of course, since
we expect nonperturbative effects to play a non-negligible
role at low $Q_T$. For comparison, Fig.~\ref{fig:cdf} also
displays the fixed-order (${\cal O}(\alpha_s)$, dotted lines, and
${\cal O}(\alpha_s^2)$, dash-dotted lines~\footnote{We have used 
subroutines of the RESBOS package of Ref.~\cite{Res} in order to 
calculate the cross section at ${\cal O}(\alpha_s^2)$.}) results for
the cross section, with their well-known divergent behavior
at small $Q_T$. 

Interestingly, the fixed-order ${\cal O}(\alpha_s)$ 
result misses the data also at {\em large} $Q_T$,
where it remains too low even if the normalization is adjusted
within the errors quoted in experiment. Joint resummation, with
the choice Eq.~(\ref{chinew}) for $\chi(\bar{N},\bar{b})$ and
the matching procedure described in Eq.~(\ref{joint:match}), adds
an important contribution to the cross section also here:
the difference $d \sigma^{\rm res}/d Q^2 d Q_T^2 -
d\sigma^{\rm exp(1)}/d Q^2 d Q_T^2$ in Eq.~(\ref{joint:match})
remains numerically significant also at large $Q_T$ and appears
to be crucial for bringing the theoretical calculation to the
data. As can be seen from Fig.~\ref{fig:cdf}, the cross section
at ${\cal O}(\alpha_s^2)$,
$d \sigma^{\rm fixed(2)} /d Q^2 d Q_T^2$, is larger
than $d \sigma^{\rm fixed(1)} /d Q^2 d Q_T^2$. 
As expected, the difference
$d \sigma^{\rm res}/d Q^2 d Q_T^2 -
d\sigma^{\rm exp(2)}/d Q^2 d Q_T^2$ is {\em smaller} than
$d \sigma^{\rm res}/d Q^2 d Q_T^2 -
d\sigma^{\rm exp(1)}/d Q^2 d Q_T^2$, so that the full cross section
in Eq.~(\ref{joint:match}) depends only little on whether matching
is performed at ${\cal O}(\alpha_s)$ or at ${\cal O}(\alpha_s^2)$.
We take this feature as an indication that our approach
for matching, Eq.~(\ref{joint:match}), is justified and reasonable even
at large $Q_T\sim Q$.

To improve the low-$Q_T$ behavior further, we introduce a
nonperturbative function as described in Eq.~(\ref{replace}).
Since most of the cross section comes from the region $Q\sim M_Z$,
we neglect the mild (logarithmic) dependence of the \nprt parameter
$g$ on $Q$. We then fit $g$ to the CDF and D0 data simultaneously,
allowing again the normalizations to vary for the two data sets.
Since $g$ should be determined from the low-$Q_T$ region, we
include only the data points with $Q_T<50$ GeV in this fit.
The results of the fit do, however, not depend much on this choice.
The optimal result is obtained for $g=0.8$ GeV$^2$ and the normalization
factors 1.069 (0.975) for CDF (D0). The $\chi^2$ for the 42 (20)
data points from CDF (D0) included in the fit is 31.3 (23.4).
Even better fits would be possible
by using a more refined nonperturbative function,
as suggested by Eq.~(\ref{explowkt}), resulting in extra parameters.
As can be seen from
the solid lines in Figs.~\ref{fig:cdf} and~\ref{fig:d0}, with these
parameters a very good agreement between the jointly resummed
cross section, matched with the fixed-order cross section at ${\cal O}
(\as)$, and the data is achieved.
We note that our nonperturbative parameter $g=0.8$ GeV$^2$ is very 
similar to that determined in Ref.~\cite{Qiu00}, where an extrapolation
of the perturbation theory result to large $b$ was made. Ref.~\cite{Qiu00}
argued that the $b$ integral in pure-$Q_T$ resummation is dominated by
the saddle point. Our method and theirs lead to somewhat smaller
values of the nonperturbative parameter than when a $b_*$ prescription
is used~\cite{ly}. 

\section{Conclusions}
The transverse momentum distribution of the Z is
by now a well-studied problem, to which
a number of successful analyses have
been applied \cite{ERV,KS,ly,Res,Qiu00}.  We have come back to
this topic because we believe that the method of
joint resummation offers additional insight on the
interplay of perturbative and nonperturbative
corrections in this, and other hadronic reactions.
Joint resummation, implemented as above,
provides a convenient definition of the
perturbative cross section at any nonzero
$Q_T$, without the introduction of additional dimensional
scales (beyond $\Lambda_{\rm QCD}$) to define
either the perturbative resummation or
the parton distributions at low scales.  

Treated this way, the jointly resummed cross section retains
its original perturbative asymptotic expansion order by
order.  It also suggests the
functional form of nonperturbative corrections.
Because perturbative and nonperturbative
components of a QCD cross section are linked
at the level of power corrections \cite{cspv,irr}, it will be
necessary, and useful, to reanalyze
the functional and phenomenological aspects
of nonperturbative corrections in this
formalism, relying on the available range
of data for the Drell-Yan mechanism.
Toward the high-energy side, a further application to
Higgs production \cite{CDG,Bal00}
at the LHC will also be of interest.
In the same spirit, we intend as well to
return to the application of joint
resummation to semi-inclusive processes such
direct photon production~\cite{lsv,lsvprl}. 

\section*{Acknowledgments}
We are grateful to A. Vogt for valuable discussions on Ref.~\cite{bv},
and for providing a version of his code for the NLO evolution of parton 
densities in which the iterative procedures for approximating the
ordered exponentials are implemented.
We are also thankful to C. Balazs, P. Nadolsky and C.P.~Yuan for helpful 
communications on the use of the RESBOS code~\cite{Res},
and we thank D.~de Florian, E.~Laenen and J.-W.~Qiu for useful discussions.
The work of G.S.\ was supported in part by the National Science Foundation,
grants PHY9722101 and PHY0098527. 
W.V.\ is grateful to RIKEN, Brookhaven National Laboratory and the U.S.
Department of Energy (contract number DE-AC02-98CH10886) for
providing the facilities essential for the completion of this work.
A.K.\ thanks the U.S. Department of Energy (contract number DE-AC02-98CH10886)
for support.

\begin{appendix}

\section{Some useful formulas}
For Z boson production one has in Eq.~(\ref{crsec}) 
the standard tree-level cross sections (see~\cite{pdb}):
\ba
\sigma_{a}^{(0)} &=& \frac{4\pi^2 \alpha^2}{9\tau S^2} \,
\hat{e}_a^2 \nonumber \\
\hat{e}_a^2 &=& e_a^2 - 2 e_a v_l v_a \kappa \frac{Q^2 (Q^2-M_Z^2)}
{( Q^2-M_Z^2)^2+M_Z^2 \Gamma_Z^2} \nonumber \\
&&+(a_l^2+v_l^2) (a_a^2+v_a^2) \kappa^2 \frac{Q^4}{( Q^2-M_Z^2)^2+M_Z^2
\Gamma_Z^2} \nonumber \\
\kappa &=& \frac{\sqrt{2}G_F M_Z^2}{4 \pi\alpha} \nonumber \\[2mm]
a_l &=& -\frac{1}{2} \, \;\;\;, \;\;\;\;\; v_l = -\frac{1}{2}+2 \sin^2\theta_W
\; ,\nonumber \\
a_a &=& T_a^3 \, \;\;\;, \;\;\;\;\; v_a = T_a^3 - 2 \,e_a \sin^2\theta_W \;.
\ea

\section{Using $x$-space parton distributions}
\label{app:xpdfs}
The above formulas are directly applicable if the parton
densities, including their evolution, are treated in Mellin
moment space. In this context, it is convenient to use
the evolution code of~\cite{grv} which is set up in
moment space. In practical applications, however, one may prefer
to be more flexible concerning the choice of parton densities
and be able to make direct use of any ($x$-space) parameterization
on the market~\cite{cteq,mrst}. One way of achieving this was
presented in Ref.~\cite{cmnt}. Here we propose a new simple
method. Let us first rewrite Eqs.~(\ref{crsec}), (\ref{cpdf}) as
\be
\label{nspace}
     \frac{d\sigma^{\rm res}}{dQ^2\,dQ_T^2}
     =  \frac{1}{2\pi i}\,\int_{C_N}\, dN \,\tau^{-N}\;
      \sum_{i,j} \;f_{i/H} (N ,\mu_F)
\; f_{j/H} (N ,\mu_F)\; \hat{\sigma}_{ij}^{\rm res} (N,Q,Q_T,\mu,\mu_F) \; ,
\ee
which is obtained after performing the $d^2b$ integration
in Eq.~(\ref{crsec}). The $\hat{\sigma}_{ij}^{\rm res}$ are
then resummed partonic cross sections, differential in $Q_T$.

The inverse-moment expression in Eq.~(\ref{nspace}) is of course
identical to an $x$-space convolution of the parton densities with
the resummed cross section~:
\be
\frac{d\sigma^{\rm res}}{dQ^2\,dQ_T^2}\;
\equiv \;\sum_{i,j}\;
\int_{\tau}^{\infty} \; \frac{dz}{z} \; \int_{\tau/z}^1 \; \frac{dy}{y}\;
\tilde{f}_i(y,\mu_F)\; \tilde{f}_j\left(\frac{\tau}{yz},\mu_F \right)\;
\tilde{\sigma}_{ij}^{\rm res} (z,Q,Q_T,\mu,\mu_F) \; ,
\label{conv}
\ee
where $z=Q^2/\hat{s}$, the $\tilde{f}_i(x,\mu_F)$ are the $x$-space
parton densities, and
$\tilde{\sigma}_{ij}^{\rm res} (z,Q,Q_T,\mu,\mu_F)$
is given by the inverse Mellin transform of the moments
$\hat{\sigma}_{ij}^{\rm res} (N,Q,Q_T,\mu,\mu_F)$,
\be
\tilde{\sigma}_{ij}^{\rm res} (z,Q,Q_T,\mu,\mu_F)\; = \;\frac{1}{2\pi i}\,
\int_{C_N}\, dN \,z^{-N}\; \hat{\sigma}_{ij}^{\rm res} (N,Q,Q_T,\mu,\mu_F)\; .
\ee
As was pointed out in~\cite{cmnt}, and as is indicated by the upper
limit $\infty$ in Eq.~(\ref{conv}), the function
$\tilde{\sigma}_{ij}^{\rm res} (z,Q,Q_T,\mu,\mu_F)$ defined 
in the ``minimal'' prescription  is non-vanishing
also at $z>1$ due to the presence of the Landau pole to the right
of the $N$ space contour (see Fig.~\ref{Ncont}), even though it
rapidly decreases with increasing $z$. At $z>1$, the angle $\phi$ of
the $N$-space contour has to be decreased to below $\pi/2$ to obtain
a convergent result.

The right-hand side of
Eq.~(\ref{conv}) in principle allows for using $x$-space parton distributions.
However, a problem arises from the fact that the resummed cross
section is highly singular~\cite{cs} at $z\to 1$ (even though regularized
in terms of plus-distributions), which makes the convolution
with the parton densities numerically very tedious~\cite{cmnt}.
A convenient way of eliminating this problem is to trivially rewrite
Eq.~(\ref{nspace}) as
\be
\label{nspace1}
     \frac{d\sigma^{\rm res}}{dQ^2\,dQ_T^2}
     =  \frac{1}{2\pi i}\,\int_{C_N}\, dN \,\tau^{-N}\;
      \sum_{i,j} \;\Big[ (N-1)\,f_{i/H} (N ,\mu_F)\Big]
\; \Big[ (N-1)\, f_{j/H} (N ,\mu_F)\Big]
\; \frac{\hat{\sigma}_{ij}^{\rm res} (N,Q,Q_T,\mu,\mu_F)}{(N-1)^2} \; .
\ee
The Mellin-inverse of $\hat{\sigma}_{ij}^{\rm res}
(N,Q,Q_T,\mu,\mu_F)/(N-1)^2$,
\be
\label{ndiv}
{\cal S}_{ij}^{\rm res} (z,Q,Q_T,\mu,\mu_F)\; = \;\frac{1}{2\pi i}\,
\int_{C_N}\, dN \,z^{-N}\; \frac{\hat{\sigma}_{ij}^{\rm res}
(N,Q,Q_T,\mu,\mu_F)}{(N-1)^2} \; ,
\ee
is now sufficiently well-behaved at large $z$ thanks to the extra suppression
by $1/(N-1)^2$. For the inverse of $(N-1)\, f_{i/H} (N ,\mu_F)$ one finds,
making use of the fact that the $x$-space parton densities vanish
at $x=1$:
\be
\label{pdfderiv}
\frac{1}{2\pi i}\,\int_{C_N}\, dN \,x^{-N}\;(N-1)\, f_{i/H} (N ,\mu_F)
= -\frac{d}{dx}\Big[x\,\tilde{f}_i (x ,\mu_F)\Big] \; \equiv \;
{\cal F}(x ,\mu_F)\; .
\ee
We thus arrive at
\be
\frac{d\sigma^{\rm res}}{dQ^2\,dQ_T^2}\;
\equiv \;\sum_{i,j}\;
\int_{\tau}^{\infty} \; \frac{dz}{z} \; \int_{\tau/z}^1 \; \frac{dy}{y}\;
{\cal F}_i(y,\mu_F)\; {\cal F}_j\left(\frac{\tau}{yz},\mu_F \right)\;
{\cal S}_{ij}^{\rm res} (z,Q,Q_T,\mu,\mu_F) \; ,
\ee
which has good numerical behavior. The standard
sets~\cite{grv,cteq,mrst} of the parton distributions allow
taking the first derivative numerically. Depending on the 
large-$N$ behavior of the resummed cross section in moment space,
$\hat{\sigma}_{ij}^{\rm res} (N,Q,Q_T,\mu,\mu_F)$, it may 
be necessary to divide by a higher power of $N-1$ in Eq.~(\ref{ndiv}), 
resulting in higher derivatives of the parton distributions
in Eq.~(\ref{pdfderiv}). This turns out to be the case for the
gluon-gluon initial state in inclusive Higgs production 
via $gg\to {\rm H}X$~\cite{forth}.

\end{appendix}

\end{document}